\documentclass[conference]{IEEEtran}

\usepackage{listings}

\usepackage{cite}
\usepackage{amsmath,amssymb,amsfonts}
\usepackage{algorithmic}
\usepackage{graphicx}
\usepackage{textcomp}
\usepackage{xcolor}
\def\BibTeX{{\rm B\kern-.05em{\sc i\kern-.025em b}\kern-.08emT\kern-.1667em\lower.7ex\hbox{E}\kern-.125emX}}

\usepackage[T1]{fontenc}

\usepackage{enumitem}

\usepackage{graphicx}
\usepackage{algorithm2e}
\RestyleAlgo{ruled}
\usepackage{amsthm}
\usepackage[hyphens]{url} 
\usepackage{subcaption}
\usepackage{tabularx}
\usepackage{booktabs}

\newcommand{\shir}[1]{\textit{\textcolor{blue}{[Shir]: #1}}}

\newcommand{\david}[1]{\textit{\textcolor{red}{[David]: #1}}}

\newcommand{\ignore}[1]{}

\newcommand{\sysName}{ReAct }
\newcommand{\sysNameNoSp}{ReAct}

\newcommand{\reqBF}{\emph{Requests\_BF} }
\newcommand{\reqBFNoSp}{\emph{Requests\_BF}}

\newcommand{\reqForTab}{\emph{Request\_Forwarding\_Table} }
\newcommand{\reqForTabNoSp}{\emph{Request\_Forwarding\_Table}}

\newcommand{\forwardedReqs}{\emph{Forwarded\_Requests} }
\newcommand{\forwardedReqsNoSp}{\emph{Forwarded\_Requests}}

\newcommand{\eg}{\emph{e.g.,}}
\newcommand{\ie}{\emph{i.e.,}}

\title{\sysNameNoSp: Reflection Attack Mitigation For Asymmetric Routing}

\author{
David Hay$^{1,2}$,
Mary Hogan$^{3}$,
Shir Landau Feibish$^{4}$\\[0.5em]
$^{1}$The Hebrew University of Jerusalem,
$^{2}$Princeton University,
$^{3}$Oberlin College,
$^{4}$University of Haifa\\[0.5em]
dhay@cs.huji.ac.il,
mhogan1@oberlin.edu,
shir@cs.haifa.ac.il
}

\begin{document}

\maketitle


\begin{abstract}
%
%
Amplification Reflection Distributed Denial-of-Service (AR-DDoS) attacks remain a formidable threat, exploiting stateless protocols to flood victims with illegitimate traffic. Recent advances have enabled data-plane defenses against such attacks, but existing solutions typically assume symmetric routing and are limited to a single switch. These assumptions fail in modern networks where asymmetry is common, resulting in dropped legitimate responses and persistent connectivity issues. 
This paper presents \sysNameNoSp, an in-network defense for AR-DDoS that is robust to asymmetry. \sysName performs request-response correlation across switches using programmable data planes and a sliding-window of Bloom filters. To handle asymmetric traffic, \sysName introduces a data-plane-based request forwarding mechanism, enabling switches to validate responses even when paths differ. \sysName can automatically adapt to routing changes with minimal intervention, ensuring continued protection even in dynamic network environments. 
We implemented \sysName on both a P4 interpreter and NVIDIA's BlueField-3, demonstrating its applicability across multiple platforms. Evaluation results show that \sysName filters nearly all attack traffic without dropping legitimate responses-even under high-volume attacks and asymmetry. Compared to state-of-the-art approaches, \sysName achieves significantly lower false positives. To our knowledge, \sysName is the first data-plane AR-DDoS defense that supports dynamic, cross-switch collaboration, making it uniquely suitable for deployment in networks with asymmetry.

\end{abstract}





\section{Introduction}

\ignore{
Outline for intro: 

\begin{itemize}
    \item Amplification attack detection in the face of asymmetric routing
    \item Describe amplification attacks + main example DNS
    \item Existing solutions: 0) are either course grained (dida) and measure only volume thus do not allow mitigation 1) use of approximate data structures such as CBF might provide false positives if the attack may alter the structure (jaqen) - meaning, if the structure is altered by the traffic (which is possibly malicious) attack traffic may cause  2) Assume a single measurement point and assume symmetric routing  - does not allow taking action in the data plane (jaqen, poseidon, etc?)
    \item To summarize vulnerabilities of amplification attacks: 1) birthday paradox if id is too small. 2) Asymmetric routing
    \item Need structure that depends only on non-malicious traffic - present ReAct.
    \item 2) need to handle asymmetric routing if taking action in the data plane - present ReAct with asymmetric routing. 
    
\end{itemize}
}

Distributed Denial of Service (DDoS) attacks have been a longstanding threat and continue to be a destructive force on the Internet~\cite{CloudflareDDoSReport}. These attacks become even more destructive when the attack traffic is seemingly legitimate, as in the case of Amplified Reflection DDoS (AR-DDoS) attacks. 
In a nutshell, AR-DDoS attacks exploit vulnerable servers on the internet, turning them into \emph{reflectors} that overwhelm the targeted victim with excessive traffic; in most cases the reflectors also \emph{amplify} the traffic, as the amount of traffic they send to the victim is much larger than the traffic they receive from the attacker(s). Perhaps the most infamous such attack is a DNS AR-DDoS attack~\cite{Paxon01}, in which an attacker sends numerous DNS requests on behalf of the targeted victim, which is then spammed by DNS responses it did not ask for. 

AR-DDoS attacks, in general, work on protocols that are built on top of connection-less communication (primarily UDP), in which a client issues a \emph{request} to the \emph{server} and waits for a \emph{response}. Servers send the response to the client based on the source IP address of the request. In addition, a \emph{transaction ID} is added to each request, and echoed on the corresponding response, so the client can match the response to its request. We note that while we use the terminology taken from the DNS protocol, this \emph{transaction-based} mechanism appears in other connection-less protocols that can be exploited for AR-DDoS attacks. For example, in Network Time Protocol (NTP), an NTP server echos the \emph{reference timestamp} of an NTP request, thus this field can be used as a transaction ID. 

Recent advancements in programmable networks allow new functionalities to be performed inside the data plane. However, despite the potential of programmable networks to support such  additional functionalities, the limited resources and processing capabilities of programmable devices pose significant challenges. Nonetheless, several solutions have been proposed for detecting and mitigating volumetric DDoS attacks~\cite{Douligeris04} and specifically AR-DDoS attacks~\cite{Paxon01} in the data plane. 

Systems such as Poseidon~\cite{zhangposeidon}, Jaqen~\cite{liu2021jaqen}, and DIDA~\cite{dida}, provide a solution for detecting and mitigating AR-DDoS using programmable switches. They rely on
different request and response counting techniques and drop responses according to these counters. Yet all of these solutions make the underlying assumption that routing is \emph{symmetric} and may significantly hinder legitimate traffic if the requests and responses do not pass through the same vantage points. 
However, the route of requests from the client to the server may not be the same as the route taken by the corresponding response. This may be due to redundancy and high availability, cost efficiency, load balancing, failures, and change of network conditions. 
This is a significant challenge when trying to detect malicious AR-DDoS attack traffic inside the network, since it requires some form of collaboration between different devices in the network. A recent study~\cite{IPD} shows that such dynamics occur often within a network, and therefore, handling asymmetric routing is crucial for providing a robust defense mechanism.

In this paper, we present a system  for mitigating amplification attacks in the data plane, for both \emph{symmetric and asymmetric} routing patterns.  We make the following contributions: 
\begin{itemize}[leftmargin=*]
    \item We design \textbf{\sysNameNoSp} (REflection AttaCk deTection),  a solution that joins legitimate requests with the corresponding responses, within the data plane, whether they traverse the same switch or not. This allows \sysName to identify legitimate traffic, so that it may block attack traffic while allowing legitimate requests to be answered.  
    \item We implement and evaluate  \sysName on the Lucid interpreter~\cite{lucid} for P4 targets and the NVIDIA BlueField-3~\cite{nvidiaBF3}.
    \item 
    For the symmetric case, we show that \sysName does not drop any legitimate responses, and can filter out most of the 
    attack traffic (exact figures depend on user-defined parameters). Importantly, unlike previous approaches, \sysNameNoSp's performance depends solely on the legitimate request rate, and remains unaffected by the volume or pattern of the attack. 
    \item 
    For the asymmetric traffic, 
    we show that \sysName blocks attacks as effectively as in the symmetric case, and incurs minimal coordination overhead. We show that this holds true for varying ratios of symmetric to asymmetric traffic.
\end{itemize}


We structure the paper as follows: We provide background on AR-DDoS attacks and discuss the limitations of existing defenses in \S\ref{sec:background}. In \S\ref{sec:design} we present the detailed design of \sysNameNoSp, and its operation under symmetric and asymmetric routing.
 Our implementation both on programmable switches and smartNICs is described in \S\ref{sec:implementation}, and \S\ref{sec:evaluation} presents our evaluation results;  \S\ref{sec:additionalProtocols} discusses potential solutions to address the limitations of \sysNameNoSp. Finally, \S\ref{sec:related} reviews related work, and \S\ref{sec:conclusion} concludes with a discussion of future directions.

\section{Background}\label{sec:background}
We provide a brief description of AR-DDoS attacks and how they are currently handled in the data plane. 

\subsection{Amplification Reflection DDoS Attacks}

Amplification-Reflection DDoS attacks (AR-DDoS) exploit the inherent asymmetry in communication protocols, where a small request generates a disproportionately large response. 
These attacks often use third-party servers or devices to amplify traffic to the target.
The attacker spoofs the source IP address, causing the response to be sent to the victim.  
%
%
%
Such attacks usually work on connection-less protocols, typically done over UDP. Since there is no pre-established connection and thus no connection identifier, in order to match the response to the request a transaction ID is often used, though not always. 
We categorize AR-DDoS attacks based on the characteristics of the identifier 
used in the protocol:
\begin{enumerate}[leftmargin=*]
    \item Fixed transaction ID in request, response and 
    retransmissions. 
    Examples of such protocols include DNS requests and responses with the same `Transaction ID' field in the payload; Memcached, in which request and corresponding response have the same `Request ID' field; and CLDAP/LDAP which uses the same `MessageID' field. 
    \item Fixed transaction ID in request, response but \emph{not} in the case of retransmissions. For example, in NTP transactions, the request sends the Origin Time-Stamp (TS). This value will be used to populate the Transmit TS in the response. However, if no response is received and retransmission is required, and a new Origin (and Transmit) TS will be used.
    \item Transactions \emph{without} transaction ID. 
    Examples of such protocols include CharGen and SSDP M-SEARCH in which requests do not include \emph{any} transaction ID.
\end{enumerate}



Traditional mitigation methods fall into two categories: software-based and hardware-based solutions.
Software-based solutions, such as Web Application Firewalls (WAFs)~\cite{booth2015network}, are typically deployed using multiple virtual machines (VMs) within a network. Incoming traffic is routed through these VMs for inspection and filtering. However, this setup introduces certain limitations. Each VM typically supports throughput between 10 and 100 Gbps, meaning dozens or even hundreds of VMs may be required to handle traffic at terabit-per-second (Tbps) scales. This can lead to increased latency, higher operational complexity, and scalability challenges. Despite these drawbacks, software-based defenses offer significant flexibility: during large-scale attacks, administrators can quickly spin up additional VMs to scale mitigation capacity as needed.

Hardware-based solutions, such as traffic scrubbing centers~\cite{collaborativeDetectARDDoS}, are purpose-built systems capable of processing very high volumes of data traffic—often in the multi-terabit range. These systems are effective for defending against large Distributed Denial of Service (DDoS) attacks but come with notable downsides. They require expensive proprietary hardware, and their architecture is generally inflexible, making it difficult to adapt or customize mitigation strategies quickly.

\ignore{
\subsubsection{Existing mitigation techniques.}\hfill\\

What is the main goal of existing techniques?  What is their basic functionality?

Two types of existing mitigation systems:
\begin{itemize}
    \item Software appliance (e.g., WAF) - set up VMs in network and divert traffic to go through them. Problems: 1) Adds latency (micro-milli-seconds). 2) Each VM can only handle O(10-100Gbps) so we need many VMs to scale the solution. Benefits - very flexible - if a new attack is seen just create the needed VMs and mitigate the attack. 
    \item Hardware appliances (e.g., traffic scrubbing center) - Problems: 1) Very expensive proprietary hardware 2) Hardware is not flexible and does not enable programmablity. 
    Benefits - can handle very large amounts of traffic. 
\end{itemize}

These solutions have significant issues. Instead, 
\sysName gets the best of both worlds - flexible and scalable but limited resources so only limited functionality is possible. This what makes programmable switches a great alternative for both performance and flexibility. 

}

\subsection{Handling AR-DDoS in the Data Plane}
\ignore{
Recent advancements in programmable networks allow new functionalities to be performed inside the data plane. However, despite the potential of programmable networks to support additional such functionalities, the limited resources and processing capabilities of programmable devices pose significant challenges. Nonetheless, several solutions have been proposed for detecting and mitigating volumetric DDoS attacks~\cite{TODO} and specifically AR-DDoS attacks~\cite{TODO} in the data plane. 

Systems such as Poseidon~\cite{TODO} and Jaqen~\cite{TODO} provide a solution for detecting and mitigating a wide variety of attacks including AR-DDoS using programmable switches. \shir{add description}
}

Programmable networks enable new solutions that can be performed right in the data plane, as packets are traversing the network, and several solutions for addressing AR-DDoS attacks in the data plane have been  proposed.

\textbf{The case for handling asymmetric routing. }
Existing techniques for detecting and mitigating network attacks in the data plane and specifically reflection attacks, often focus on a single switch solution. In Jaqen~\cite{liu2021jaqen}, for example, requests are maintained in a counting Bloom filter (CBF). When a response is received, if the corresponding request is found (\ie~the relevant counters are all greater than $0$), it is deleted (\ie~the counters are decremented) and the response continues to its destination. If the request is not found the response is dropped. In DIDA~\cite{dida}, counters are maintained based on source and destination IP addresses. Specifically, switches count outgoing requests from the client to each server, and incoming responses from the server to each client. 
If more responses than requests are received, the system assumes that the responses are attacks and drops responses accordingly. 
In both of these solutions, if routing is \emph{asymmetric}, the requests and responses may not go through the same switch. In this case, legitimate responses could be dropped as they will not be matched with the corresponding response. In fact, it is possible that even upon retransmission the response will be dropped, thus inadvertently disconnecting the source of the requests from the service, which could cause the source significant harm. For example, if all of the DNS requests  from a certain client are routed through a different path than the responses, no DNS responses will reach the client, thus essentially leaving the client without access.  

\textbf{Using attack traffic to modify the structure. } 
Both of these solutions are directly affected by the rate of the attack. 
For example, Jaqen uses a CBF where the transaction IDs are deleted upon receiving a corresponding response. This may cause the system to incorrectly classify legitimate responses as attacks, if the attack responses (even by chance) cause the structure to decrement an index to $0$ inadvertently. We evaluate such a scenario in \S\ref{sec:eval_cbf} and show that the rate of these misclassifications may be significant.

\section{\sysName Design}
\label{sec:design}
In AR-DDoS attacks, many messages are sent to a victim device or entity, which contain seemingly legitimate responses to requests made by the victim. Yet, in an attack, the majority of  these requests were \emph{not} made by the victim. 
In order to identify these unwanted responses before they reach the victim, \sysName must be able to match each response to the correlating request, if indeed such a request was made. \sysName will maintain a succinct representation of the requests seen, so that these may be joined with the responses received.

We first describe how \sysName can be used to detect and mitigate AR-DDOS attacks in a single switch when routing is symmetric. We then go on to describe how \sysName can also be used to mitigate such attacks when routing is asymmetric.  

In this section we will focus on AR-DDoS attacks in protocols with a fixed transaction ID in both the initial request and response and also in retransmissions (\eg~DNS, Memcached). For ease of explanation we will describe our system using the DNS protocol as an example, though additional protocols with the same behavior can be handled in the same manner. We discuss protocols in which transaction IDs are modified in request retransmission in \S\ref{sec:additionalProtocols}.



\begin{figure}
    \centering
    \includegraphics[trim = 0 195 0 30, clip, width=0.99\linewidth]{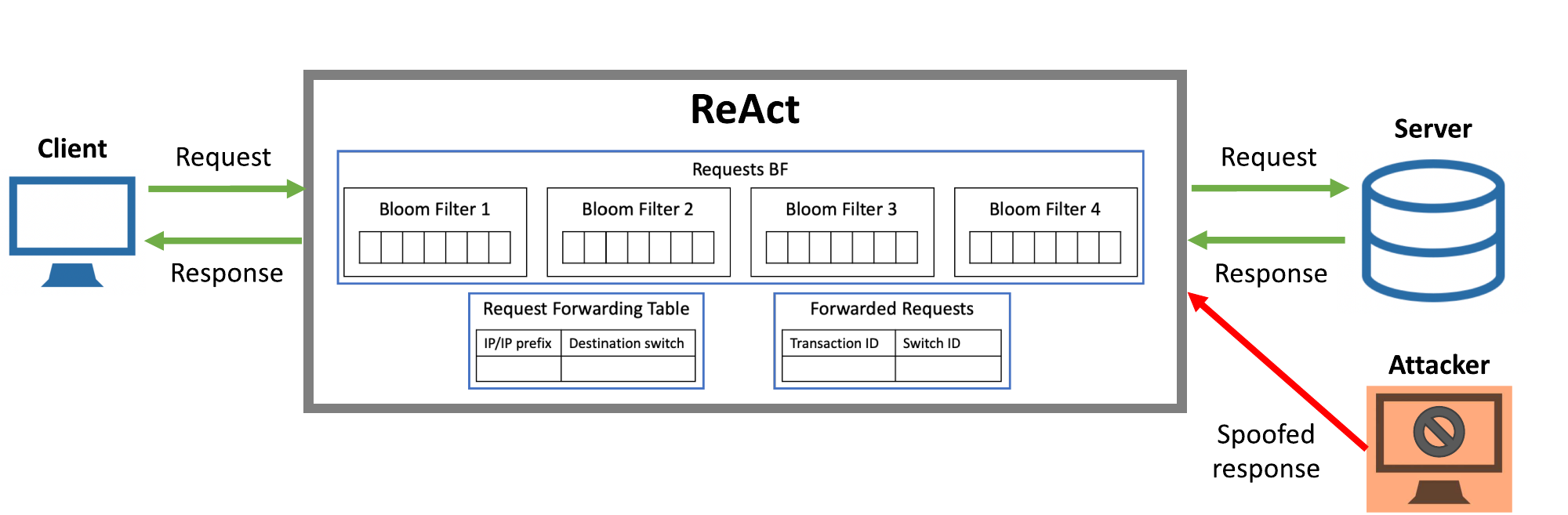}
    \caption{\sysName in Symmetric Routing}
    \label{fig:reactSymmetric}
\end{figure}


\ignore{
Network setting: Request and response must go through the same point
\begin{itemize}
    \item 
        Data Structures: Bloom filters.
    \item 
        Describe key - How to code ``transaction IDs''
    \item 
        Cleaning-up (getting rid of old transactions) - sliding windows
    \item
        Implications of false positives and negatives. Numbers (maybe in case of small sliding window)
    \item 
        Dealing with IP fragmentation in the responses. (we check only first fragment, all the rest will be forwarded assuming client will drop due to missing fragments) 
    \item 
        P4 implementation (Switch and NIC? Separate solutions?)
\end{itemize}
}

\textbf{\sysName Overview. }
In order to verify that responses are legitimate, \sysName keeps track of the requests that it sees, and uses this information to match each response with its respective request. 
Ideally, it would be best if each of the requests could be maintained in the data plane with all of the relevant information, but unfortunately, the limited resources of the data plane do not allow this.  

Instead, \sysName maintains an approximate representation of the requests that have recently been seen. When a response is received, it is first checked to see if it matches one of the existing requests. If such a request exists, the response continues on its way to its destination; otherwise, it is dropped. 

If the requests and responses go through the same vantage point (i.e. routing is symmetric), this join process is straightforward. If, however, routing is asymmetric, the request and response may not traverse the same switches. In this case, the  
switch receiving the request will identify the switch receiving the response and forwards the request to the relevant switch. 

Our approach works under the following assumptions: (i) each request traverses at least one programmable switch, which we call the \emph{upstream switch}; (ii) each response goes through at least one programmable switch, which we call the \emph{downstream switch}; (iii) the control planes of the upstream and downstream switches implement the same logic and one can send packets from one switch to the other. Furthermore, we assume this communication's latency to be smaller than the latency between the switch(es) and the DNS server.
(iv) If a transaction fails (\eg~a DNS request does not receive a response), the client retries by sending another request with the same ID. This is a common practice in most DNS client implementations, as it allows the client to keep track of pending requests and delayed responses. We note that assumption (iv) does not hold for all protocols. For example, in NTP  a new transaction ID is generated for every retry; we discuss this limitation further in \S\ref{sec:additionalProtocols}. 


The system handles asymmetric response paths by coordinating between upstream and downstream switches. When a request is received, the upstream switch checks if it is a retransmission. In this case, it broadcasts a copy of the request to all switches in order to discover where responses are arriving. The downstream switch that receives the response informs the upstream switch about the correct downstream path for future responses. This mechanism ensures proper request-response matching even when paths are asymmetric.

\begin{figure}
    \centering
    \includegraphics[trim = 0 172 0 12, clip,width=0.86\linewidth]{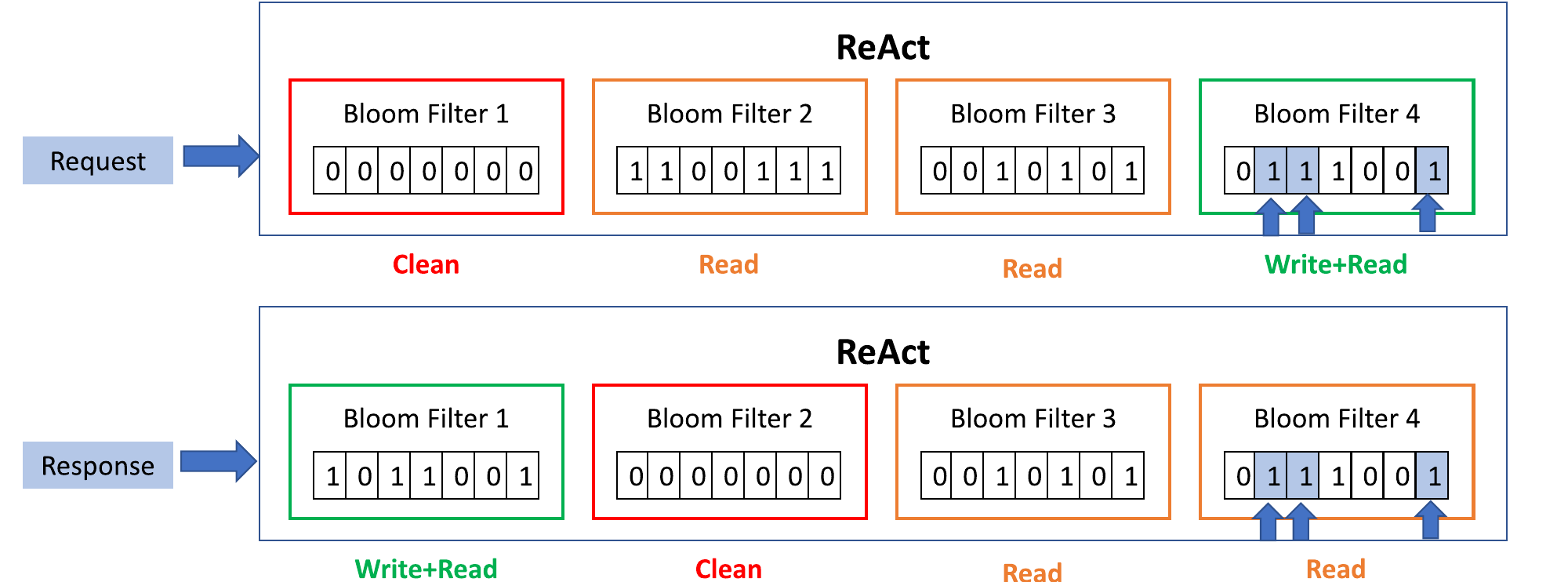}
    \caption{\sysNameNoSp's sliding window structure}
    \label{fig:react}
\end{figure}

\textbf{\sysName Components. }
\sysName maintains three main parts: 
\begin{itemize}[leftmargin=*]
    \item \reqBFNoSp. 
\sysName uses a series of Bloom filters (BFs)~\cite{bloom-filter} to keep a succinct summary of the requests. The BFs will be used in a `sliding window' manner to enable both the insertion of new requests and the eviction of outdated requests.
By using sliding windows, \sysName avoids the need to use response traffic to actively remove requests from the structure as done in Jaqen~\cite{liu2021jaqen}, making \sysName more robust to volumetric attacks. 


\item \reqForTabNoSp. This table indicates which switch or switches might receive responses from a given source or IP prefix. This mapping will be used to forward requests received at the switch.  That is, when a request is received at a switch, \sysName uses this table to identify if this request needs to be cloned and forwarded, and if so, to which switch(es). This may be implemented as a key-value store or as a match-action table. 

\item \forwardedReqsNoSp. This table maintains information regarding requests that have been forwarded to a given switch. When \sysName is trying to identify if the response for a certain request is received at some switch in the network, it will forward the request to that switch so that it may correlate between this request and the relevant response. 

\end{itemize}

\subsection{\sysName for Symmetric Routing}\label{sec:symmetric}

If routing is symmetric, \sysName need only keep track of requests and responses received at the same switch using the \reqBFNoSp, as shown in Fig.~\ref{fig:reactSymmetric}. 
When receiving a request, the request key, composed of the transaction ID along with the \emph{source} IP, is hashed into the Bloom filter. When a response is received, the response key, composed of the transaction ID and the \emph{destination} IP, is hashed to see if there is a correlating request. In a legitimate request-response scenario, the request key should be the same as the response key. 

Over time, the BF may become overcrowded, 
thus increasing false positives. Furthermore, responses are expected to be received in a timely manner, and if the response is not received before the request timeout, the client may issue a new request, so 
\sysName must evict old requests from the structure. This will be enabled with a series of BFs which will be used as `sliding windows'. 
In each interval of time, one BF will be written to, at least one BF will be read from and one BF will be cleaned.
As shown in Fig.~\ref{fig:react}, when the request comes in, it is added to the BF that is set at that time to be the one used for writes. The BFs will switch roles in a round-robin manner, such that the BF that was most recently written to is used only for reads and the ``oldest" BF will be cleaned in the next interval, and the cleaned BF will be written to. 
Note that due  to the memory access restrictions, bulks of memory cannot be accessed; thus each index in the memory that needs to be cleaned must be accessed and reset individually. 
This can be done with a helper packet that is recirculated multiple times so that it can clean out the entire structure, as done in prior work (~\cite{conquest,MIDST}). 

Two main user-defined parameters directly impact system performance. 
The first is the \emph{size of each Bloom filter} in \reqBFNoSp. Due to collisions, BFs may incur false positives. The probability for false positives increases as the size of the structure decreases.\footnote{The false positive rate also depends on the number of hash functions used, which is defined by the user.}  
The second is the \emph{interval length}, which determines the rate at which the role of windows will be interchanged and thus the amount of time that a request will be maintained by the system. False positives increase as the number of items inserted into the filter is increased. \sysName may simultaneously handle multiple types of attacks, and therefore, the interval length needs to be selected so that it allows 
sufficient time for legitimate responses to come back and still be able to find the correlating request. 
Assume the sliding window has a size of $\tau$ and utilizes $b$ BFs.
Note that \sysName should check both the BFs that are being read as well as the one being written to, in which case 
a `check' operation for $x$ at time $t$ will always return true if $x$ was inserted into the data structure during the interval $(t-\tau(b-2), t]$. 

\subsection{\sysName for Asymmetric Routing}\label{sec:asymmetric}

\begin{figure*}
    \centering
    \includegraphics[width=0.78\linewidth]{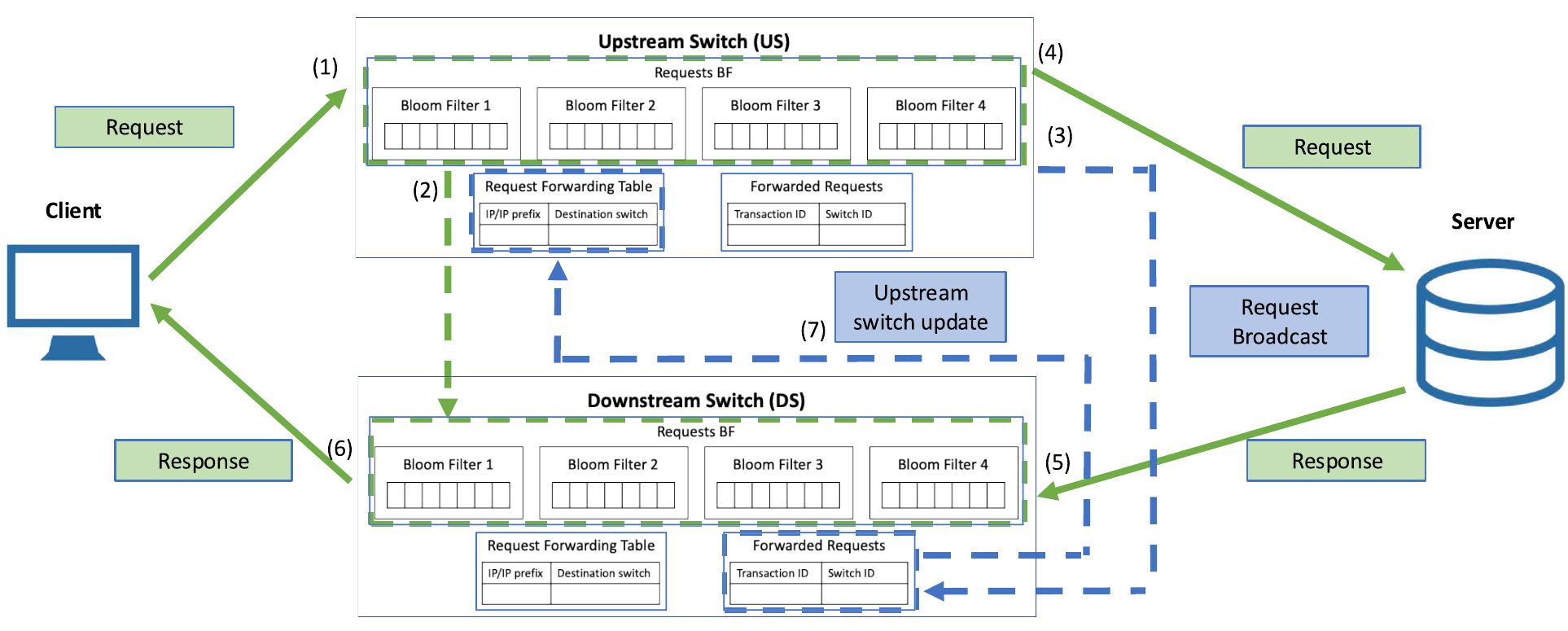}
    \caption{\sysName in Asymmetric Routing}
    \label{fig:reactASymmetricExample}
\end{figure*}

We now relax the underlying traffic symmetry assumption 
and devise a framework and protocol to mitigate reflection attacks even when asymmetric routing is deployed. 
As seen in Fig.~\ref{fig:reactASymmetricExample}, the request is sent through switch 1, which is the upstream switch. The response is sent through switch 2, which is the downstream switch. Notice that we do not make any assumptions about the path taken by the attack packets. 

\textbf{Dealing with asymmetry when routing is known. }
We begin by outlining a strategy for a simplified scenario in which the paths for requests and responses are distinct but known to the network controller. 
%
Since routing is known to the controller apriori, 
the controller sets a rule for each source or IP prefix in the \reqForTab of the upstream switch, which indicates that the requests from these sources should be forwarded to another downstream switch.  
Upon receiving a request in the upstream switch, if the request matches one of the forwarding rules, the switch generates a duplicate of the DNS request with a specific mark and sends this duplicate to the specified downstream switch. 
This downstream switch processes these marked DNS requests as if they were standard DNS requests. That is \sysName inserts the request into the relevant Bloom filter, yet it \emph{refrains} from forwarding these requests to the DNS server. 

We highlight that our approach involves duplicating DNS \emph{requests} rather than DNS \emph{responses}, and we assume those requests originate from legitimate internal clients. This distinction is key because the traffic generated during reflection attacks consists of responses. Consequently, by adopting this method, the additional DNS traffic generated by our solution is proportional to the volume of \emph{legitimate} DNS traffic, and remains so even under a volumetric reflection attack. We note that a DoS attack from duplicated attacks is only possible if an attack within the network engages in IP spoofing, which is typically mitigated via mechanisms like egress filtering.

\textbf{Dealing with asymmetry when routing is not known. }
We now remove the assumption that routing is known ahead of time, and thus the upstream switch will need to identify the downstream switch. 
 We assume that each switch has a unique ID, denoted \texttt{switch\_id} drawn from the set of IDs $S$. 

 \ignore{
 Each switch holds the following data structures: A primary bloom filter \reqBF 
 and $|S|$ auxiliary Bloom filters for  \texttt{forwarded\_requests[id]}, one Bloom filter for each $id\in S$. As described above, each switch also holds a  \emph{request\_forwarding\_table} used to decide whether to duplicate a request and forward it to other switches. We note that having $S$ Bloom filters might cause significant scalability issues when $S$ is large, thus in \S~\ref{sec:forward_rules} we present an algorithm that uses only a configurable number of Bloom filters between $1$ to $|S|$. 

\shir{this has some scalability issues if the number of switches is big. We could group switches in a single BF - giving a tradeoff between the number of BFs and the amount of communication. Lets also try to think of better options...}\david{I think we should first present a ``clean'' solution and then try to optimize it. For example, in most cases, I don't expect a specific upstream switch to have that many downstream switches, so most of these bloom filters will be empty and shouldn't be considered. }
}

For each request received at an upstream switch, \sysName needs to determine if the downstream switch of the response will be a \emph{different} switch, so that it may send the request to the relevant switch as needed. 
Upon receiving a \emph{request}, \sysName checks if the request is a \emph{retransmission}. Such detection can be done by a regular `find' operation on \reqBF (namely, checking all BFs that are not in the `clean' state), or by refraining from checking the BF that is currently being written to, thus assuming the retransmission period is between $\tau$ and $\tau(b-2)$, where $\tau$ is the interval length of the sliding window and $b$ is the number of BFs. Notice that if the retransmission period 
always exceeds $\tau$, the latter choice reduces the number of broadcasts due to misclassification. 

In any case, \sysName logs the request by inserting it into the relevant BF of the \reqBFNoSp. 
If the request is not a retransmission, \sysName checks the \reqForTab to see if there is already a rule for the source of the request. If such a rule is found, \sysName will duplicate the request. The original request will be sent on its way to the server, and the duplicate request will be marked and forwarded to the relevant downstream switch. The downstream switch will log the message to its \reqBF (as if it is a regular request) but will refrain from forwarding it.  

If the upstream switch detects that the request is retransmitted, it will mark it and broadcast it to all other switches.
When a switch receives a broadcasted request, it will log it in its \reqBF as well as in its  \forwardedReqs key-value table, along with the 
switch ID that broadcasted the message. 
This process is described below: 
\vspace{-0.3em}

\begin{lstlisting}[linewidth=0.95\columnwidth,breaklines=true, basicstyle=\footnotesize\ttfamily,numbers=left, firstnumber=1, escapechar=\%]
Upon receiving request r=(src,dst,req_id,mark):
   if mark is null:  /* upstream switch */ 
        if r not in Requests_BF:
            insert r to Requests_BF 
            forward r (to DNS server)
            res = apply_match(src, Request_Forwarding_Table)
            if res is not null:
                forward r to res with additional_mark 
                    (forward, switch_id)
        else: /* r is retry */
            insert r to Requests_BF /* to make sure it is not deleted prematurely */
            broadcast r with additional mark
                (broadcast,switch_id) to all switches                 
    else if mark is forward:
        insert r to Requests_BF
    else if mark is (broadcast,id):
        insert r to Requests_BF;
        Forwarded_Requests[req_id].append(id); 
        /* key is req_id, value is a list with switch ids the requests were broadcasted from */
\end{lstlisting}
\vspace{-0.3em}

Copies of requests and responses that are forwarded between the switches are marked and forwarded accordingly. Requests can either be marked by \texttt{forward} or \texttt{broadcast}, along with the  \texttt{switch\_id}, where 
\texttt{forward} indicates that the forwarding is done to a specific downstream switch(es), that previously handled corresponding responses for those requests; 
\texttt{broadcast} indicates that this is a request that is being sent for the second time, meaning that the original request did not receive a response (possibly due to asymmetric routing), and therefore, the switch tries to identify where (and if) the response is handled, by broadcasting to all relevant switches in the network. Note that broadcast is typically done while bootstrapping the system or when routes are changed. 

When a response is received by the downstream switch, the switch checks to see if a correlating request is found and handles the response accordingly. Additionally, it checks to see if a correlating request is found in \forwardedReqs. If so, the response is duplicated and sent to the upstream switch that is indicated in the table (with mark \texttt{forwarding\_rule}), so that the relevant forwarding rule may be added to the upstream switch(es). Notice that this happens only for responses that match previously broadcasted requests (implying that there were no forwarding rules in place). Furthermore, there is no reliable transport mechanism (\eg~an acknowledgment packet) on these response copies; if the copy is lost, forwarding rules are not processed by the upstream switch(es). In such a (rare) case, the upstream switch has no choice but to broadcast the next request following the same asymmetric routing and wait for its corresponding forwarding rule. 

The process of rule insertion may require controller assistance if the \reqForTab is a match-action table (\ie~a TCAM table), in which case the downstream switches will update the controller.  If the \reqForTab is a key-value store, the upstream switch may insert the relevant key and switch ID into the table from within the data plane. Note that if the response traverses multiple downstream switches \sysName may create a forwarding rule to a group of switches, essentially creating a multicast rule. 

This process is described below:
\vspace{-0.3em}
\begin{lstlisting}[linewidth=\columnwidth,breaklines=true, basicstyle=\footnotesize\ttfamily,numbers=left, firstnumber=1, escapechar=\%]
Upon receiving response r=(src,dst,req_id,mark):
    if mark is null:  /* downstream switch */ 
        if r is in Requests_BF
            forward r to dst /* else it is dropped */
        if forwarded_requests[req_id] is not empty
            for each id in forwarded_requests[req_id]
                forward r to id with additional mark %\hfill%(forward_rule,switch_id)
    else if mark is (forward_rule,id):
        insert match-action rule to Request_Forwarding_Table with rule.src=dst/16,%\footnote{Notice that we have arbitrarily chosen to use prefix mask /16, though any other prefix length can be selected.}%  rule.action.add(id)                    
\end{lstlisting}
\vspace{-0.3em}

Fig.~\ref{fig:reactASymmetricExample} shows an example of this process: 
Switch 1 is the upstream switch and Switch 2 is the downstream switch.
(1) When Switch 1 receives a request, \sysName will check if the same request is found in \reqBF to see if it is a retransmission and will log it by adding it to the relevant BF in \reqBFNoSp. 
If it is not retransmitted \sysName will check \reqForTabNoSp. (2) If the source is matched, it will 
be sent to at least one downstream switch. If it is a \emph{retransmission}, that has previously been logged, it needs to find out if a different switch is receiving the responses.
(3) In order to determine which downstream switch is getting the responses, \sysName will duplicate the request, mark it with \texttt{forward} and broadcast it to all other switches. (4) In any case, the switch will also send the original request on its way. 
When Switch 2 receives such a broadcast, it will log the request in \reqBF and add it to \forwardedReqs table, indicating that the request was received from Switch 1. 
(5) When Switch 2 receives a response, it will check if it is found in \reqBFNoSp. (6) If it is, it will send the response back to the client, otherwise it will be dropped. (7) It will then check if it is also found in \forwardedReqsNoSp, and if it is, it will initiate an update to the \reqForTab of the relevant upstream switch (\ie~Switch 1). 
Note that based on the implementation of the \reqForTabNoSp, this process may involve the controller.  
Subsequent requests from this source that are received at Switch 1  will be duplicated and sent to Switch 2 as indicated in step (2) above. 

\section{Implementation}
\label{sec:implementation}


\subsection{Implementation of \sysName on a Programmable Switch}
We implement \sysName in Lucid~\cite{lucid}, a high-level abstraction for P4~\cite{p4} with a C-like syntax. In Lucid, incoming packets are represented as \emph{events}, whose corresponding handlers are executed upon packet arrival.
The Lucid backend includes a compiler to P4, which we then use to compile to programmable hardware (e.g., Intel Tofino). 
Lucid also provides an interpreter to simulate a program without compiling it. The interpreter runs a network-wide simulation with multiple switches, allowing us to simulate the asymmetric case.

We choose Lucid because its abstractions significantly simplify the process of programming P4. Additionally, while Lucid currently only supports the Tofino, P4 is being increasingly supported by different programmable devices (e.g., SmartNICs~\cite{bluefieldp4, pipeleon}, FPGAs~\cite{netfpgap4, intelfpga}, XSight~\cite{XSightPR}). 

When compiled to the Tofino, our implementation of \sysName requires 11 out of the 12 total stages and uses four BFs, each with $2^{17}$ bits, which is approximately 6\% of the total available memory. We use at most 24\% of all other resources.

\subsection{Implementation of \sysName on a SmartNIC}
\label{sec:dpu_implementation}

We have implemented \sysName on the NVIDIA BlueField-3)~\cite{nvidiaBF3}, a high-performance SmartNIC that integrates both programmable hardware and software processing capabilities. Our deployment assumes a symmetric model in which all DNS requests originate from the client and all responses return to the SmartNIC. This assumption eliminates the need to handle asymmetry, and allows us to compare against existing approaches that only handle symmetric traffic.
We assume a symmetric model because SmartNICs are typically placed near the client; thus they would see both the request and response. 

BlueField-3 combines two processing domains: general-purpose ARM cores for software processing, and a programmable hardware pipeline based on the disaggregated Reconfigurable Match Tables (dRMT) model~\cite{chole2017drmt}. However, the hardware pipeline lacks general-purpose registers.
As a result, it cannot maintain the per-flow state required by \sysNameNoSp.
To address this, we adopt a hybrid design. The hardware pipeline is used to perform fast, stateless operations such as packet duplication, header manipulation, and forwarding, while the ARM cores handle stateful logic. 
Unlike P4 registers, memory-managed BFs can be dynamically allocated and replaced. We leverage this flexibility to use only two BFs per core in a sliding window structure: one for \emph{read and write} and one for \emph{read} only (cf. Fig.~\ref{fig:react}; a BF in the \emph{clean} phase is not needed). A designated control core periodically creates a new BF, updates the pointer for the corresponding core, thus switching write operations to this new BF, and frees the memory of the old one. 



\section{Evaluation}
\label{sec:evaluation}
To evaluate the performance of \sysNameNoSp, we consider the following key parameters:

\begin{itemize}[leftmargin=*]
  \item \emph{Request traffic rate $r$ (requests per second):} The average number of DNS requests sent from the client per second.
  \item \emph{Number of Bloom filters $b$ and hash functions $k$:} For Lucid/P4, the minimum number of Bloom filters is 3; for BlueField-3, it is 2 per core. In our experiments, we set $b=4$ for Lucid and $b=2$ for BlueField-3, with $k=2$ fixed throughout.
  \item \emph{Interval length $\tau$ (sliding window):} Filters are rotated every $\tau$ seconds. To handle asymmetric routing, $\tau(b-2)$ must exceed the DNS client retransmission time $T$. We note that the default retransmission time (namely, timeout) $T$ is between $1$ to $5$ seconds for most DNS clients~\cite{microsoft_dns_timeouts, man7_resolv_conf, microsoft_nslookup_timeout, linux_die_net_dig}.
  \item \emph{Total filter size $s$ (in bits):} The total memory allocated across all BFs.
\end{itemize}

These parameters define the load $\lambda = \frac{b\cdot r\cdot \tau}{s}$ on each BF, which determines its false positive rate $\varepsilon=(1-e^{-k\lambda})^k$. \sysNameNoSp's overall misclassification probability ranges from $1-(1-\varepsilon)^{b-2}$ (just after a swap) to $1-(1-\varepsilon)^{b-1}$ (just before).\footnote{For the BlueField-3 implementation, the respective probabilities are $1-(1-\varepsilon)^{b-1}$ and $1-(1-\varepsilon)^b$.}

\sysName may suffer from two types of misclassifications:
\begin{itemize}[leftmargin=*]
  \item \emph{False negatives (FNs) on attack traffic:} Malicious responses are not dropped.
  \item \emph{False broadcasts (asymmetric case):} Misclassified DNS requests falsely appear as retransmissions, triggering unnecessary broadcasts. If $T>\tau$, we can avoid checking the most recent BF, thus reducing the false broadcasts rate to at most $1-(1-x)^{b-2}$.
\end{itemize}
Legitimate responses are never dropped in the symmetric case, and may only be dropped before the first retransmission in the asymmetric case.

\subsection{Experiments the BlueField-3 Implementation}

We evaluate \sysName on the BlueField-3 DPU.
The client sends $r$ DNS requests per second with random transaction IDs and source ports, querying the same domain. A local BIND9 DNS server~\cite{isc_bind9} responds from its cache. To simulate network conditions, a delay $d$ with 10 ms jitter is introduced. An attacker injects $a$ forged responses per second with random transaction IDs and destination ports. We log (legitimate) responses from the BIND9 server and compare them with those received by the client to identify false negatives. We also verify that no false positives occur. 
Experiments run for one minute on 14 ARM cores, each with two filters of size $2^{13}$ bits, totaling $s=2 \cdot 14 \cdot 2^{13} = 229{,}376$ bits. We note that our implementation can handle up to 8.07 million malicious responses \emph{per core} without dropping any legitimate responses.  

\begin{figure}[t]
    \centering
    \subfloat[False negative rate vs. request rate $r$]{
        \includegraphics[width=0.68\linewidth]{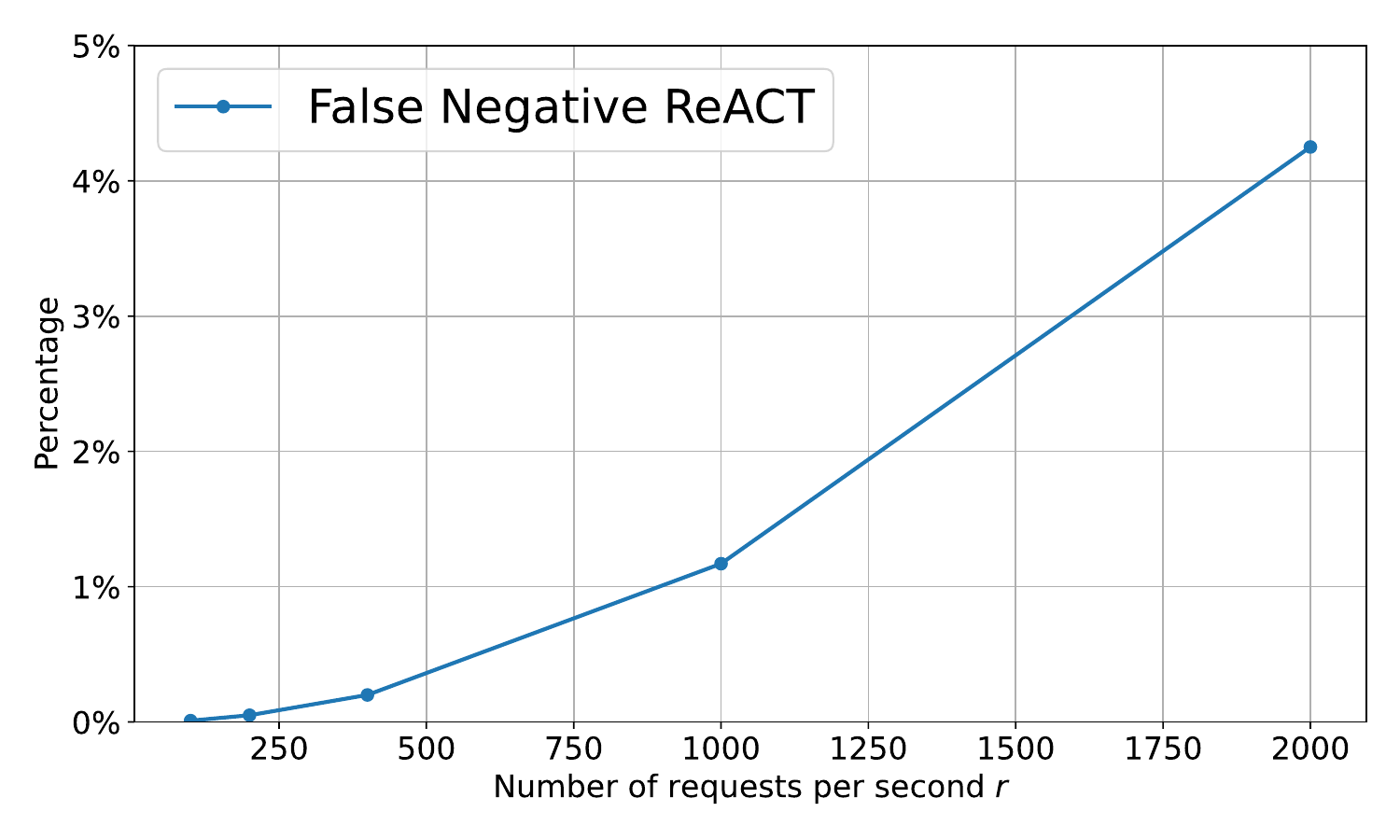}
        \label{fig:fn_requests}
    }
    \\
    \subfloat[False negative rate  vs. interval length $\tau$]{
        \includegraphics[width=0.68\linewidth]{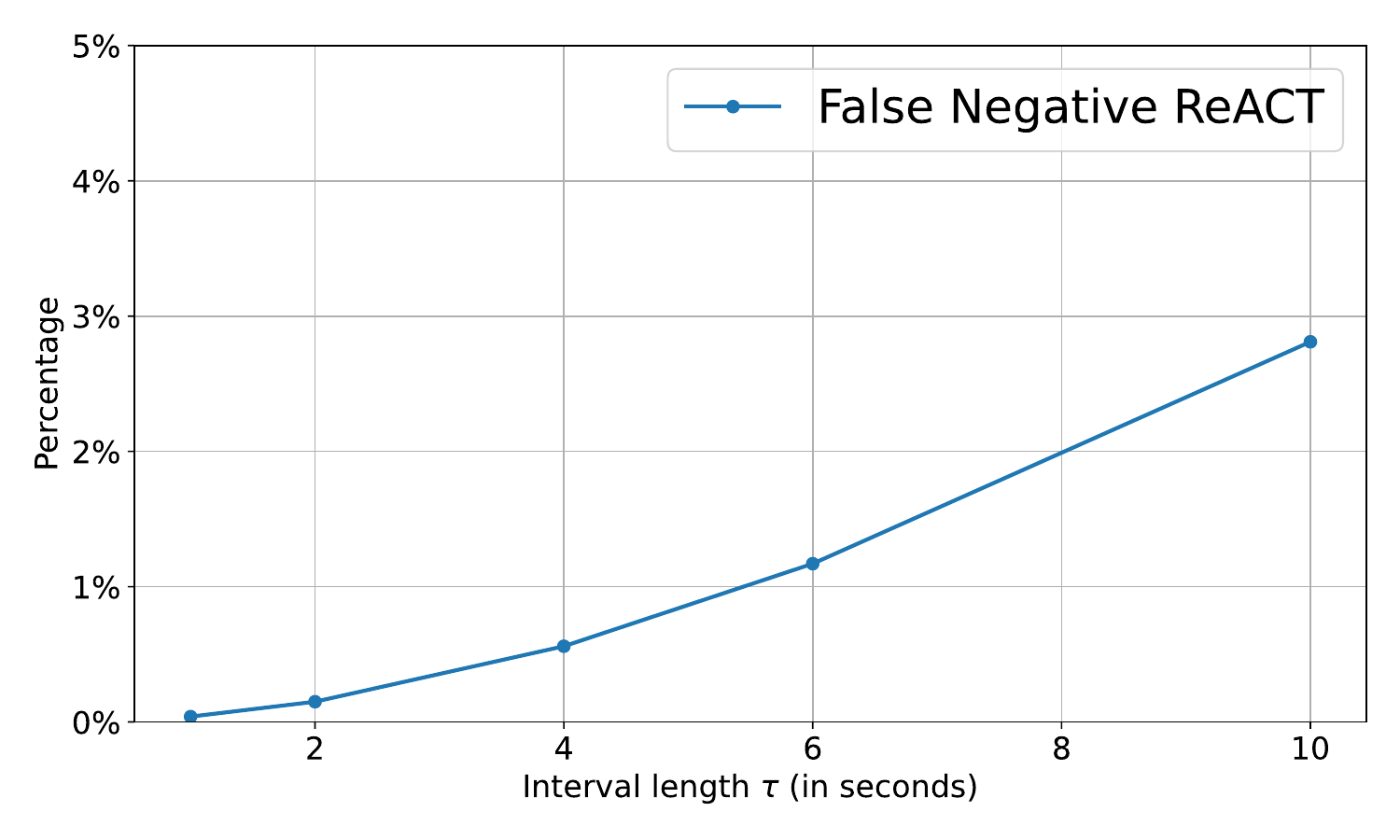}
        \label{fig:fn_intervals}
    }
    \caption{Sensitivity of \sysName to the request rate $r$ and interval length $\tau$. The false negative rate captures the attack traffic that is misclassified and forwarded to the client.}
    \label{fig:fn_combined}
\end{figure}

\textbf{Parameter Sensitivity.}
\label{sec:eval_fn}
We analyze how varying $r$ or $\tau$ affects \sysNameNoSp's false negative rate, as both increase the load $\lambda$. Fig.~\ref{fig:fn_combined} shows results: in (a), $\tau=6$ seconds is fixed and $r$ is varied; in (b), $r=100$ requests per second is fixed and $\tau$ is varied. In both, $a=100000$ requests per second and $d=100$ms. These latter parameters do not impact \sysNameNoSp's performance (as discussed above). 
Since the BlueField-3 is typically deployed near the client, it does not need to handle asymmetry (see Section~\ref{sec:dpu_implementation}). As a result, the interval length $\tau$ can be set much smaller than the retransmission timeout $T$, and tuned instead to the network delay $d$. This implies that, in practice, very low false negative rates can be achieved. 

\textbf{Comparison with Counting Bloom Filters (CBFs).}
\label{sec:eval_cbf}
To compare with Jaqen~\cite{liu2021jaqen}, we implemented a CBF on BlueField-3. Each core uses a single CBF of size $s/b$ with 1-byte counters. Sliding windows are not used, as they are not needed.

CBFs may introduce false positives (legitimate traffic being dropped): if an attack response matches all relevant counters and decrements them before the legitimate response arrives, the latter may be dropped. During the delay period $d$, roughly $a \cdot d$ attack packets are sent, each attempting to collide with the relevant counters, implying that as this product increases, so does the false positive rate.

\begin{figure}
\centering
    \subfloat[\sysName vs. CBF under varying attack ratios $a/r$,]{
    \includegraphics[width=0.70\linewidth]{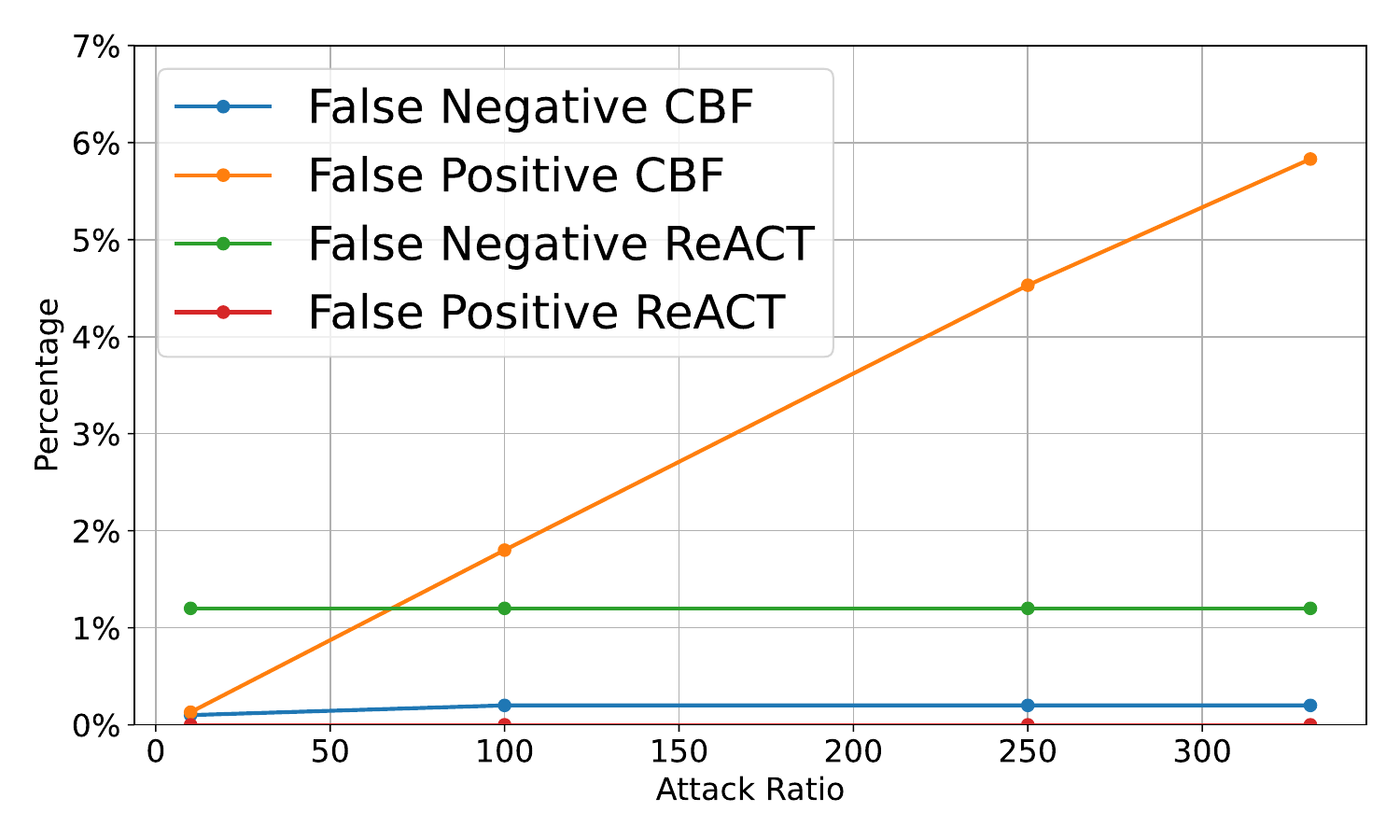}
    \label{fig:CBFattack}
    }
    \\
    \subfloat[\sysName vs. CBF under varying delay $d$.]{
    \includegraphics[width=0.70\linewidth]{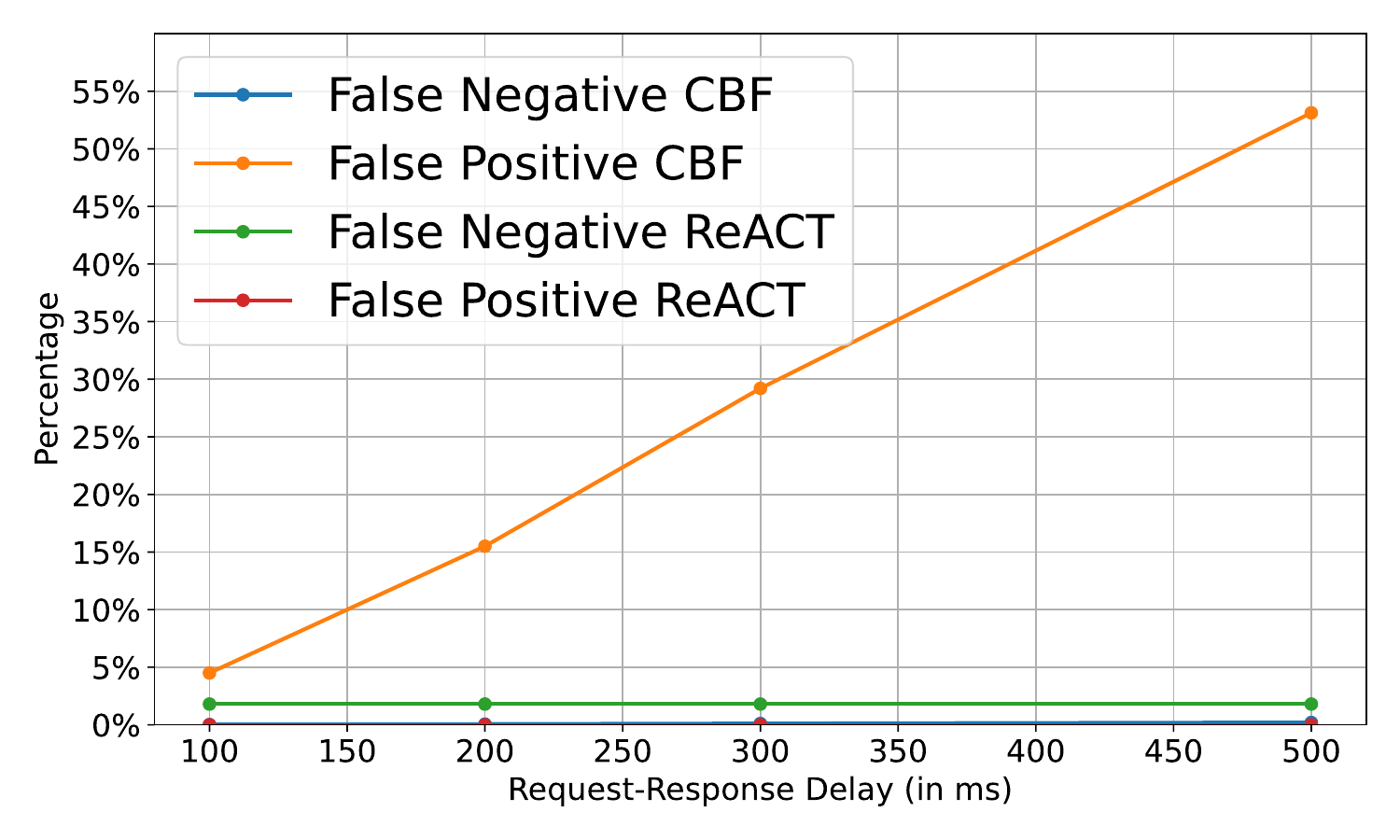}
    \label{fig:CBFdelay}
    }
    \caption{Comparison between \sysName and CBF. 
   } 
\end{figure}

We first compare \sysName and CBF under varying attack ratios $a/r$. Fig.~\ref{fig:CBFattack} shows both false negatives and false positives rates with fixed $\tau=6$ seconds, $r=1000$ requests per second, $d=100$ ms, and $s=229{,}376$ bits. As expected, \sysName remains unaffected by attack volume, while CBF suffers increasing false positives as $a/r$ grows. 
Fig.~\ref{fig:CBFdelay} fixes $a/r=250$ (namely, $250{,}000$ attack responses per second and $1{,}000$ legitimate responses per seconds) 
and varies $d$. Again, \sysName remains unaffected, while CBF suffers increasing false positives as the delay $d$ grows. Notably, at $d=500$ ms, CBF drops over half of legitimate responses.

It is important to note that even moderate false positive rates render CBFs \emph{unsuitable} for the asymmetric case, as each false positive triggers a retransmission, which in turn leads to a broadcast. A large number of such broadcasts may significantly increase network congestion and fill up data structures in other switches, making them ineffective.

\subsection{Experiments with the Lucid implementation}

We also evaluate \sysName on simulated programmable switches with Lucid. Our simulation has two switches, where requests always go through the upstream switch. Symmetric responses also go through the upstream switch, while asymmetric responses go through the downstream switch. Attack responses may go through either switch.
The switch receives $r$ DNS requests per second with random transaction IDs, source ports, and client source IPs. In our experiments, we set $r=7000$ requests per second, chosen such that it is large enough for the size of our BFs ($2^{17}$ bits), but small enough to be scalable for our simulator (the Lucid interpreter processes approximately 100000 packets per second).

Similar to the Bluefield-3 setup, packets have a delay $d=100$ms with 10 ms jitter, and an attacker injects $a=r*10$ forged responses per second. We generate corresponding responses in our simulation, and measure both the requests that are retransmitted and the responses that are dropped. 

Experiments simulate 30 seconds of requests, using four filters of $2^{17}$ bits, and an interval length $\tau =$ 4 seconds. We set the size of the \forwardedReqs table to 2048 entries, with each entry expiring after 1 second. This size worked well in our experiments, but can easily be increased, especially if there are no other programs requiring memory on the switch.

To evaluate the efficacy of \sysName in the asymmetric case, we vary the percentage of asymmetric traffic, and we see that \sysName is blocks almost all attack traffic, while incurring minimal overhead (after the initial bootstrapping). We measure the false negative rate in Table~\ref{fig:fn_asym}, and see that \sysName performs similarly in the asymmetric case as it does in the symmetric case, blocking over 97\% of attack traffic on average, regardless of how much traffic is asymmetric. We note that, as shown in \S\ref{sec:eval_fn}, this holds for any attack traffic rate.

\begin{table}[tb]
\centering
\small
\begin{tabular}{cc}
\toprule
\textbf{\% Symmetric} & \textbf{Avg false negative rate} \\
\midrule
\textbf{30\%} & 2.5\% \\ 
\textbf{50\%} & 2.5\% \\
\textbf{70\%} & 2.1\% \\
\bottomrule
\end{tabular}
\caption{\sysName false negative rate for varying ratios of symmetric to asymmetric traffic.}
\label{fig:fn_asym}
\end{table}

\textbf{\sysName Overhead.} We analyze the overhead incurred by retransmitting and broadcasting requests. Recall that false positive rate captures the number of dropped legitimate responses. Initially, we see in Fig.~\ref{fig:asym_fp} that \sysName has a high false positive rate; essentially all of the asymmetric responses are dropped. Once the DNS clients timeout (after 5 seconds), requests are retransmitted. We see a spike in broadcasts at 5s in Fig.~\ref{fig:asym_broadcasts}, which corresponds to the initial retransmissions. Fig.~\ref{fig:asym_broadcasts} shows the number of requests broadcasted in comparison to the total number of requests sent (including retransmissions).
After requests in this bootstrapping phase are broadcasted, \sysName can correctly forward DNS responses, and install table entries in the \reqForTab in the upstream switch. 

Once \sysName stabilized (at 10 seconds in our simulation), the average broadcast rate was 7\%, 4\%, 2\% for 30\%, 50\%, 70\% symmetric traffic, respectively. This corresponds to approximately 230, 120, 53 broadcasted requests. These broadcasts happen because after stabilization, our upstream switch sees asymmetrically routed requests from new prefixes, that are not yet captured in the \reqForTabNoSp.



\ignore{
\begin{figure}[t]
    \centering
    \subfloat[False positive rate for varying ratios of symmetric to asymmetric traffic. False positive rate captures the legitimate traffic that is misclassified and dropped. ]{
        \includegraphics[width=0.63\linewidth]{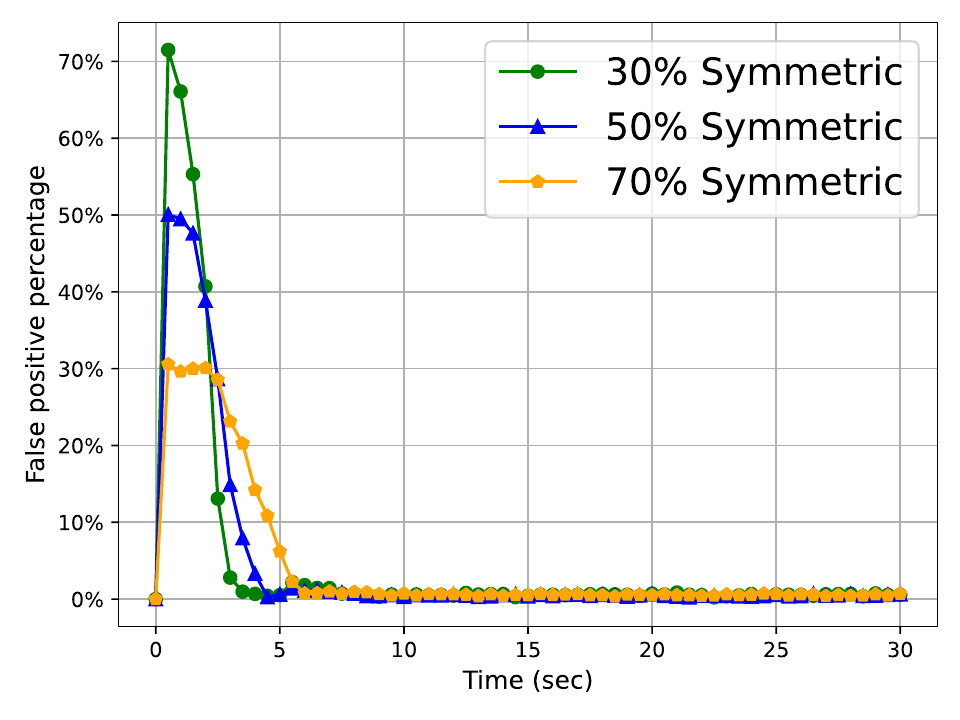}
        \label{fig:asym_fp}
    }
    \\
    \subfloat[ The broadcast rate measures the number of broadcasts compared to total requests sent.]{
        \includegraphics[width=0.69\linewidth]{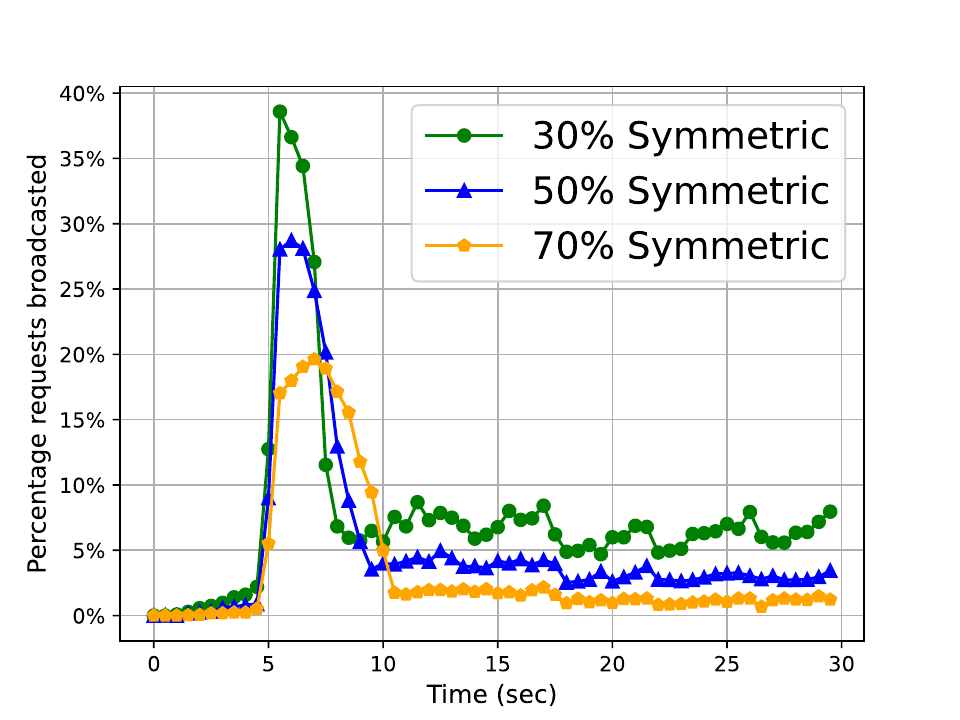}
        \label{fig:asym_broadcasts}
    }
    \caption{Overhead of \sysName for asymmetric traffic.}
    \label{fig:asym_combined}
\end{figure}
}

\begin{figure}[t]
    \begin{subfigure}{0.48\textwidth}
    \centering
    \includegraphics[trim=0 0 0 0, clip, scale=0.32]{Figures/falsepositivesNoType3Fonts.pdf} 
    \caption{False positive rate (misclassified legitimate traffic).}
    \label{fig:asym_fp}
    \end{subfigure}
    \begin{subfigure}{0.48\textwidth}
    \centering
    \includegraphics [trim=0 0 30 0, clip, scale=0.35]{Figures/broadcastratesNoType3Fonts.pdf} 
    \caption{Broadcast rate (broadcasts as a fraction of total requests).}
    \label{fig:asym_broadcasts}
    \end{subfigure}

\caption{Overhead of \sysName for varying ratios of symmetric to asymmetric traffic. }
\label{fig:asym_combined}
\end{figure}

\section{Limitations}\label{sec:additionalProtocols}

In protocols without a fixed transaction ID (\eg~NTP) our symmetric solution will work, but we'll need to solve this issue for asymmetric routing.

AR-DDoS attacks can be performed over various protocols, including: DNS, NTP, Memcached, etc. In the hope of providing a robust mechanism for detection and mitigation of AR-DDoS attacks we would like to provide a solution that can handle all of the different protocols simultaneously using a single data structure. This will also require maintaining the requests for the appropriate time interval, such that requests are kept in the structure long enough for the relevant response to return but not too long to overload the structure. 

The basic functionality of \sysName is a join operation between each response and its respective request. 
 To do so, \sysName must first determine a way to correlate the requests and responses. 
Clearly, the source of the request (and destination of the response) are important in this identification, yet are not sufficient. 
While certain protocols such as DNS provide a transaction ID that is found in both the request and response, in other protocols this may not hold. For example, in NTP the ID is based on the timestamp of the request and therefore retransmissions 
will be based on a different timestamp.

\textbf{Dealing with asymmetry in NTP. }
In NTP, each transaction is uniquely identified by a timestamp (namely, the Origin Timestamp T1), which \emph{is not} reused for retry attempts. This hinders the ability to leverage retries for detecting asymmetry.

To address this, we broadcast requests whenever there is no corresponding entry in the forwarding table (as opposed to waiting for a retransmission). To prevent the network from being overwhelmed, especially for requests that already correspond to symmetric paths, we introduce an auxiliary table \texttt{forward\_table[switch\_id]} that tracks symmetric routing by logging responses received at the upstream switch, for which a correlating request was found (when the responses are received where requests are being tracked). Hence, we only broadcast a single request for a source sub-network.

Notice this solution introduces a challenge not present in the previous setting: the risk of erroneously categorizing the responses of a source IP address as going through the wrong downstream switch, due to false positives of the Bloom filter or a change of routes.  Such an error could result in the persistent rejection of all NTP responses until the time window elapsed, and retries will not initiate broadcasts to correct that error. While this problem can be solved by monitoring unanswered requests and purging corresponding forwarding rules, practical implementations of NTP clients issue another request only after several minutes and often apply exponential backoffs (DNS retries are issued within seconds). This implies that by setting the forwarding rules expiry time to be smaller than that, these requests will be broadcasted. Yet, forwarding rules reduce the number of broadcasts as they are installed per sub-network. 
Dealing with protocols without transaction IDs requires a different approach and we leave this to future work. 

\textbf{Eliminating the need to wait for retransmissions. }
In addition to the above method, 
we can alternatively identify asymmetric routing on the downstream traffic by initiating broadcasts of unmatched incoming responses. However, as mentioned before, this, by itself, might cause an amplification attack. Thus we need to ensure that such broadcasts are done only in \emph{peace time}; namely, when the network has enough resources to handle these broadcasts. This is done by keeping track of the ratio between the number of unmatched responses and the total number of responses at the switch and broadcast only if it is below a certain configurable threshold. Otherwise, the response is dropped as before. Notice that since responses are not always broadcasted, this mechanism should work in conjunction with the previous mechanisms that broadcast requests (and works even if the network is under attack). 
\section{Related Work}
\label{sec:related}

\textbf{Traditional defense mechanisms. } 
Host-based mitigations, such as SYN cookies~\cite{syncookie}, and middlebox appliances, like Arbor TMS and NSFocus ADS, mitigate floods by inspecting stateful traffic off-path. Similarly, export protocols like NetFlow~\cite{netflow} and IPFIX~\cite{ipfix} rely on packet sampling and defer in-depth analysis to centralized collectors. While this is effective in some contexts, these solutions often suffer from high latency and ususally require costly hardware.  They are also ineffective against spoofed traffic, particularly in the presence of asymmetric routing. In contrast, \sysName operates directly within the forwarding ASIC or SmartNIC, enabling line-rate filtering with \emph{orders-of-magnitude} lower memory usage compared to off-path collectors.

\textbf{Defense mechanisms in programmable networks. }
Defense systems based on SDN or NFV~\cite{bohatei, senss, SmithS18} have been introduced to mitigate DDoS attacks by leveraging available resources to dynamically allocate mitigation capacity. However, relying solely on software-based solutions does not scale well and legitimate traffic may be rerouted through multiple mitigation virtual machines, leading to increased latency.


Network monitoring and telemetry using programmable switches (\cite{univmon,HashPipe,marple,sonata, nitrosketch}) focus on generic flow statistics, heavy-hitter detection, or query-driven telemetry, but do not join requests and responses and therefore cannot determine if a response is unsolicited. A line of work uses counts of DNS (or other similar protocols) requests and responses entirely inside one data-plane device~\cite{zhangposeidon,liu2021jaqen,dida}.  
These works use counting Bloom filters, per–prefix counters, or hierarchical sketches to detect an imbalance of responses and requests.  
However, the request and matching response must traverse the \emph{same} switch, meaning that all three systems assume \emph{symmetric routing}, and legitimate responses are dropped whenever the forward and reverse paths diverge. Recently we have also seen DDoS mitigation solutions in both FPGAs~\cite{pham2017fpga, ddos_spectral_filter, fpga_survey} and NPUs~\cite{netBouncer}.

\textbf{Recent multi-device defenses.}
Very recent systems such as \textsc{DNSGuard}~\cite{duan2025dnsguard}, \textsc{HELP4DNS}~\cite{sahin2025help4dns} and \textsc{DAmpADF}~\cite{dai2024dampadf} extend counting techniques with machine-learning features or filters that may be updated. Nonetheless, these solutions still rely on observing both directions at the same vantage point, hence the papers do not discuss or evaluate path asymmetry.

\section{Conclusion}
\label{sec:conclusion}

We present \sysNameNoSp, a framework for mitigating reflection attacks within the data plane for both symmetric and asymmetric traffic. We believe this framework can be useful for performing asymmetric joins. For example, the solution could be extended to handle attacks on additional protocols, such as response-based DDoS attacks in connection-based communication (e.g., SYN floods and TCP reset attacks). Effectively mitigating these attacks requires tracking corresponding connection identifiers (e.g., 4-tuples of source and destination IPs and ports) to verify whether a given response is associated with a legitimate request, which is challenging due to memory and scalability constraints, especially in high-throughput environments with millions of simultaneous connections. Nonetheless, in scenarios where the duration for which state must be maintained is very short, such as when responses are expected within a tight time window or traffic patterns are predictable, this requirement may be practical. We leave this to future work.


The authors have provided public access to their code  at~\cite{code}.

\textbf{Acknowledgments}
This work was partially supported by the Fraunhofer Institute and the Israel Science Foundation (No. 980/21). We thank Abed Kahtib for helpful discussions.

\sloppy
\bibliographystyle{IEEEtran}
\bibliography{sample-base,software,references}


\end{document}